\documentclass[a4paper,11pt]{article}
\pdfoutput=1 

\usepackage{jheppub} 

\usepackage[T1]{fontenc} 

\title{\boldmath Quantum phenomenological gravitational dynamics: A general view from thermodynamics of spacetime}

\author[a]{A. Alonso-Serrano}
\author[b]{M. Li\v{s}ka,}

\affiliation[a]{Max-Planck-Institut f\"ur Gravitationsphysik (Albert-Einstein-Institut), ,\\Am M\"{u}hlenberg 1, 14476 Potsdam, Germany}
\affiliation[b]{Institute of Theoretical Physics, Faculty of Mathematics and Physics, Charles University,
\\V Hole\v{s}ovi\v{c}k\'{a}ch 2, 180 00 Prague 8, Czech Republic}

\emailAdd{ana.alonso.serrano@aei.mpg.de}
\emailAdd{liska.mk@seznam.cz}

\abstract{In this work we derive general quantum phenomenological equations of gravitational dynamics and analyse its features. The derivation uses the formalism developed in thermodynamics of spacetime and introduces low energy quantum gravity modifications to it. Quantum gravity effects are considered via modification of Bekenstein entropy by an extra logarithmic term in the area. This modification is predicted by several approaches to quantum gravity, including loop quantum gravity, string theory, AdS/CFT correspondence and generalised uncertainty principle phenomenology, giving our result a general character. The derived equations generalise classical equations of motion of unimodular gravity, instead of the ones of general relativity, and they contain at most second derivatives of the metric. We provide two independent derivations of the equations based on thermodynamics of local causal diamonds. First one uses Jacobson's maximal vacuum entanglement hypothesis, the second one Clausius entropy flux. Furthermore, we consider questions of diffeomorphism and local Lorentz invariance of the resulting dynamics and discuss its application to a simple cosmological model, finding a resolution of the classical singularity.}


\begin{document}
\maketitle
\flushbottom

\section{Introduction}

The search for the elusive theory of quantum gravity has been a fundamental pillar of research in the last decades, and although there are promising candidates, none of them has yet provided a consistent final theory. In this context, phenomenology of quantum gravity has acquired interest as an approach to extract information about the possible dynamical effects of quantum gravity theories in the low energy regime~\cite{Kempf:1994,Garay:1994en,Smolin:2016,Chakraborty:2019}. Its results have mainly focused on the understanding of physics close to classical singularities~\cite{Adler:2001,Awad:2014,Alonso:2018}. The limitations of these models lie, on one side, in the extreme simplicity of the studied cases, whose results also cannot be directly extrapolated to extract more general features of the theory. On the other side, models attached to a particular theory are not well suited to provide any general constraints for the final theory of quantum gravity.

The motivation of this paper came precisely from the search for general phenomenological effects of quantum gravity. The development of complete effective dynamics, that can also be particularised to specific solutions, would allow one to explore global properties of spacetime in the presence of quantum gravity effects, which should be recovered by the candidate theories of quantum gravity. As we will see later, thermodynamics not only provides tools to deal with the emergence of gravitational dynamics, but also its predictions are common for most of the candidate theories.

Thermodynamics of spacetime has been revealed as a very useful tool to understand gravitational mechanics. Since the seminal developments of black hole thermodynamics, it has been attempted to extend this framework to general spacetimes in order to understand the relation between thermodynamics and geometry~\cite{Wald:1993,Jacobson:2003}. Along these lines, the derivation of Einstein equations from thermodynamics of (local virtual) Rindler horizons~\cite{Jacobson:1995ab} puts on the table a new idea for the emergence of classical gravitational dynamics just from thermodynamic concepts applied to gravity. A concept of matter entropy crossing the horizon is defined via the Clausius relation, $\text{d}S=\delta Q/T$, where $\delta Q$ is the matter-energy flux crossing the horizon and $T$ its associated Unruh temperature. Equilibrium condition between this entropy and entanglement entropy associated with the horizon, then, implies Einstein equations; when the entanglement entropy follows the Bekenstein formula and one considers local horizons constructed in every point of spacetime. This basic derivation has been further developed later in the literature, polishing and generalising the original arguments~\cite{Eling:2006,Padmanabhan:2010,Chirco:2010,Jacobson:2012,Baccetti:2013ica,Jacobson:2015,Bueno:2017,Svesko:2017,Svesko:2019}.

Overall, thermodynamics of spacetime implies equations of motion of general relativity (GR) and several modified theories of gravity~\cite{Eling:2006,Padmanabhan:2010,Jacobson:2012,Bueno:2017,Svesko:2017,Svesko:2019}, and it also provides insight into situations where the source of gravity are quantum fields rather than classical matter~\cite{Jacobson:2015,Svesko:2019}. Therefore, it appears natural to go one step further and employ thermodynamic methods for gaining some insight into the low energy gravitational dynamics of quantum gravity. If we modify thermodynamics of spacetime to include quantum gravity effects, the corresponding emergent gravitational dynamics can be expected to encode the low energy limit of quantum gravity and it will provide quantum phenomenological equations of gravitational dynamics that are independent of any specific model of quantum gravity. Moreover, it generalises the idea of finding equations of motion for spacetime from thermodynamics, extending the discussion about the interface between gravity and thermodynamics.

Here, we will focus on two recent derivations, extensively reviewed in a previous paper of the authors~\cite{Alonso-Serrano-Liska} (see this reference also for a more detailed historical review of thermodynamics of spacetime). The first one obtains the dynamics from thermodynamic equilibrium of geodesic local causal diamonds (GLCD), by performing a simultaneous variation of the entanglement entropy associated with the horizon of GLCD and the entanglement entropy of the matter present inside it~\cite{Jacobson:1995ab}. The second derivation utilises as the starting point an expression for the Clausius entropy flux across any null bifurcate surface~\cite{Baccetti:2013ica} applied to the specific case of causal diamonds. Equations of motion are then found by comparing the decrease of Clausius entropy of the matter fields with the corresponding change of entanglement entropy of the horizon~\cite{Alonso-Serrano-Liska}. As it was shown there, both derivations give rise to the same gravitational equations of motion in the semiclassical case. The interest of considering both derivations lies, on one side, in checking the consistency of results, and, on the other side, in studying the correspondence among the different concepts of entropy. The most relevant results of the analysis of these derivations were the recovery of unimodular gravity (UG) instead of GR and the semiclassical equivalence of Clausius and entanglement entropy~\cite{Alonso-Serrano-Liska}. Note that UG implies the same classical dynamics as GR, but the quantisation of both theories might differ. Then, the introduction of quantum gravity effects could provide a preferred direction towards one or another theory of gravity. Likewise, the entropy equivalence reported in~\cite{Alonso-Serrano-Liska} might break due to quantum gravity corrections. These questions provide further motivation for extending thermodynamics of spacetime into the realm of phenomenological quantum gravity.

In this paper, in order to provide robust and sufficiently general results, we consider the leading order quantum correction to the Bekenstein entropy formula, namely a logarithmic term in the horizon area. The presence of a term of that form is predicted by various approaches to quantum gravity, such as loop quantum gravity (LQG)~\cite{Kaul:2000,Meissner:2004}, string theory~\cite{Banerjee:2011,Sen:2013} and AdS/CFT correspondence~\cite{Carlip:2000,Fareghbal:2018}. Logarithmic corrections also arise from phenomenological models such as generalised uncertainty principle (GUP)~\cite{Adler:2001}, which adds an extra non-commutative term to the well-known Heisenberg uncertainty principle, due to introduction of a minimal length into the theory. GUP is an effective prediction implied by string theory, path integral gravity and several model-independent thought experiments~\cite{Garay:1994en}. Moreover, the entanglement entropy associated with closed causal horizons (even virtual ones) also develops logarithmic corrections of the same form as Bekenstein entropy. In fact, it has been broadly proposed in the literature to interpret Bekenstein entropy as the entanglement entropy~\cite{Sorkin:1986,Srednicki:1993,Solodukhin:2011}. In summary, the presence of logarithmic corrections to horizon's entropy appears to be rather universal result. Therefore, the modifications of gravitational dynamics we obtain are relevant for many different approaches to quantum gravity.

The paper is organised as follows. In section~\ref{log_entropy}, we review calculations of the logarithmic corrections to the entanglement entropy associated with local causal horizons and their connection with quantum gravity effects. In section~\ref{modified_equations}, we first introduce a geometric tool we will use in the derivation, that is, geodesic local causal diamonds. Then, we use both above described thermodynamic methods~\cite{Alonso-Serrano-Liska} to derive the quantum phenomenological gravitational equations of motion. Section~\ref{interpretation} discusses diffeomorphism and local Lorentz invariance of the dynamics we obtained. Furthermore, we apply the new equations of motion to a simple cosmological model. The resulting modified Friedmann and Raychaudhuri equations suggest a possible replacement of the Big Bang singularity with a quantum bounce. Lastly, section~\ref{discussion} sums up our results and outlines possible future developments.

Throughout the paper, we work in four spacetime dimensions and use metric signature $(-,+,+,+)$. Definitions of the curvature-related quantities follow~\cite{MTW}. We use lower case Greek letters to denote abstract spacetime indices and lower case Latin letters for spatial indices with respect to a (local) Cartesian basis. Unless otherwise explicitly stated, we use the SI units.

\section{Bekenstein entropy, entanglement entropy and corrections to them}
\label{log_entropy}

In this section, we review the appearance of a logarithmic correction term in the Bekenstein entropy equation. As we will see, such a term arises due to quantum gravity effects not only in black hole entropy but also in entanglement entropy associated with observer-dependent causal horizons. In section~\ref{modified_equations}, we will show how these corrections give rise to quantum modifications of gravitational dynamics.

Bekenstein equation for black hole entropy states
\begin{equation}
\label{BH_entropy}
S_{BH}=\frac{k_B\mathcal{A}}{4l_P^2},
\end{equation}
where $\mathcal{A}$ is the area of the black hole's event horizon, $l_P=\sqrt{G\hbar/c^3}$ is the Planck length and $k_B$ is the Boltzmann constant. Bekenstein entropy is implied by the combination of semiclassical effects (driven by Hawking radiation) and the fully classical first law of black hole mechanics~\cite{Bekenstein:1973,Bardeen:1973,Hawking:1975}. Its possible microscopic interpretations draw much attention and many proposals have been put forward~\cite{Wald:2001}. One of the ideas is that Bekenstein entropy arises due to quantum entanglement between two causally separated regions~\cite{Sorkin:1986}. An observer in one region cannot access information in the other one. Since there are correlations between vacuum fluctuations in both regions, some information is inaccessible and non-zero entanglement entropy arises~\cite{Sorkin:1986}. It was shown that for a Klein-Gordon field (either massless or massive) in flat spacetime, this entropy is infinite unless an ultraviolet cutoff length is introduced~\cite{Sorkin:1986}. Then, entanglement entropy is directly proportional to the horizon area
\begin{equation}
\label{entanglement}
S=\eta\mathcal{A},
\end{equation}
where $\eta$ can, in principle, depend on the position in spacetime~\cite{Chirco:2010}. Since its introduction~\cite{Sorkin:1986}, this result has been confirmed and expanded to more general cases~\cite{Srednicki:1993,Das:2008,Solodukhin:2011}. Entanglement entropy proportional to area appears whenever a part of spacetime is inaccessible to observer measuring the entropy. Thus, it can be associated with any causal horizon, including observer-dependent ones such as the acceleration (Rindler) horizon~\cite{Sorkin:1986,Srednicki:1993,Chirco:2010}. However, calculations of entanglement entropy are unable to fix the proportionality constant $\eta$; which, to recover Bekenstein equation, should have universal value $\eta=k_B/4l_P^2$. The main issue with the identification of black hole entropy as entanglement entropy is the latter's dependence on the number of fields and their coupling to gravity~\cite{Solodukhin:2011}. Presently, there is no widely accepted resolution of these issues. Nevertheless, both problems are resolved for instance in the Sakharov's induced gravity scenario, as quantum fluctuations responsible for the entanglement entropy also induce a renormalisation of the Newton's gravitational constant in such a model~\cite{Jacobson:1994}.

It has been argued in various contexts~\cite{Solodukhin:1995,Kaul:2000,Carlip:2000,Adler:2001,Das:2002,Fareghbal:2018} that the semiclassical Bekenstein entropy should be modified when one takes into account quantum gravity effects. Then, there emerge modifications beyond the linear term in area, generalising the Bekenstein equation into
\begin{equation}
\label{modified_BH}
S_{BH,q}=\frac{k_B\mathcal{A}}{4l_P^2}+\mathcal{C}k_B\ln{\left(\frac{\mathcal{A}}{\mathcal{A}_0}\right)}+O\left(\frac{k_Bl_P^2}{\mathcal{A}}\right),
\end{equation}
where $\mathcal{C} \in \mathbb{R}$ is a dimensionless constant and $\mathcal{A}_{0}$ is a constant with dimensions of area. Values of both constants depend on the specific model. Logarithmic modifications to black hole entropy were found, e.g. in LQG~\cite{Kaul:2000, Meissner:2004}, GUP phenomenology~\cite{Adler:2001,Alonso:2018}, entanglement entropy calculations~\cite{Solodukhin:1995,Solodukhin:2011}, AdS/CFT duality~\cite{Carlip:2000,Fareghbal:2018}, string theory \cite{Banerjee:2011,Sen:2013} and in the analysis of statistical fluctuations around equilibrium~\cite{Das:2002}. 

Of the above mentioned approaches, entanglement entropy calculations are especially well applicable in our situation. Firstly, they allow us to explicitly find the logarithmic term  even for certain virtual, observer-dependent horizons. Secondly, they tell us that such corrections will not appear in general, but instead depend on the horizon's topology. For these reasons, we will proceed by reviewing the derivation of the logarithmic corrections to a spherical horizon in flat spacetime. The procedure is described in detail, e.g. in~\cite{Mann:1998,Solodukhin:2011,Solodukhin:2020}.

\subsection{Logarithmic corrections to the entanglement entropy}
\label{entanglement_log}

In 4-dimensional Minkowski spacetime consider a 2-sphere of radius $l$ at $t=0$. Then perform a Wick rotation $\tau=it$ to obtain 4-dimensional Euclidean spacetime. The resulting Euclidean metric in the spherical spatial coordinates is conformal to
\begin{equation}
\label{H2S2 metric}
\text{d}s^2=l^2\left(\text{d}\rho^2+\rho^2\text{d}\chi^2\right)+l^2\text{d}\Omega^2.
\end{equation}
This metric describes a direct product of a 2-hyperboloid $H_2$ and a 2-sphere $S_2$, both of radius $l$, where $\rho\in[0,\infty)$, $\chi\in[0,2\pi)$ are polar coordinates on the 2-hyperboloid, $\tau=\rho\sin\chi$, $r=\rho\cos\chi$ and $\text{d}^2\Omega=\sin\theta\text{d}\theta\text{d}\phi$, with $\theta\in[0,\pi)$, $\phi\in[0,2\pi)$.

Now assume that a massless, Lorentz invariant scalar field is present in the spacetime. Using the so-called replica trick, we can calculate the entanglement entropy that arises due to tracing over the sphere's interior~\cite{Mann:1998,Solodukhin:2011}
\begin{equation}
S_e=\frac{k_B \mathcal{A}}{48\pi\varepsilon^2}-\frac{k_B}{180}\ln\frac{\mathcal{A}}{4\pi\varepsilon^2}+\text{UV finite},
\end{equation}
where $\varepsilon$ is the UV cutoff length and the UV finite terms are disregarded. Considering $\varepsilon\approx l_P$, we obtain a result consistent with Bekenstein equation in the leading term, but with a negative correction logarithmic in the horizon area. The entropy has the same structure for any minimally coupled quantum field, regardless of its spin and mass~\cite{Solodukhin:2011}.

Let us note that logarithmic corrections are connected with the Euler number of the horizon surface $\Sigma$~\cite{Solodukhin:2011}, that for a 2-sphere equals $2$. However, for a plane (Rindler horizon) the Euler number is $0$. Due to this difference, a logarithmic term is present for spherical black holes as well as for virtual spherical horizons in flat spacetime, but not for Rindler horizons. Therefore, in order to exploit the logarithmic term to obtain quantum corrections to Einstein equations, we need to consider a horizon with closed spatial cross-section, in contrast with the original Jacobson's paper~\cite{Jacobson:1995ab}.

Up to this point, we have only discussed the contribution coming from a single minimally coupled scalar field. The total entropy is, of course, a sum of contributions coming from all the quantum fields present in spacetime, including gravitons. Graviton contribution also most clearly shows the connection between logarithmic corrections to entanglement entropy and quantum gravity. The classical Einstein-Hilbert gravitational action, $S=\left(1/16\pi G\right)\int R\sqrt{-g}\text{d}^4x$, yields only the leading order term in entropy. The term proportional to $\ln\varepsilon$ comes from corrections to the action that are quadratic in curvature. Finally, the term proportional to $\ln\mathcal{A}$ is connected with non-local quantum corrections to gravitational action~\cite{Solodukhin:2020}. For local actions that do not contain derivatives of Riemann tensor, entanglement entropy can be even related to Wald entropy~\cite{Wald:1993}, as both the replica trick calculation of entanglement entropy and the Wald calculation of entropy as a Noether charge amount to evaluating the same integral~\cite{Solodukhin:2011}.

As we will see, the sign of logarithmic correction term in entropy determines the physical implications of the modified equations of motion we obtain. Most notably, it indicates a clear resolution of spacetime singularity for positive signs but not for negative ones. Therefore, we provide a brief overview of signs obtained in entanglement entropy calculations, as well as in other approaches that imply logarithmic corrections to entropy.

The logarithmic term arises even in entanglement entropy of black hole horizons. However, its sign for a Schwarzschild black hole is positive~\cite{Solodukhin:1995}. On the other hand, a negative sign was obtained for extremal and near extremal Reissner-Nordstr\"{o}m black holes~\cite{Solodukhin:2011} (the sign changes at $r_{-}=2r_{+}/3$, with $r_{\pm}$ being the outer and inner horizon radius). The reason is that the Schwarzschild spacetime near the horizon is given by the product of a 2-sphere and a 2-disk, while an extremal spherically symmetric black hole near the horizon has the geometry of the product of a 2-sphere and a 2-hyperboloid. Thus, for a spherical extremal black hole, the calculations leading to entropy of a sphere in flat spacetime can be reproduced with minimal modifications~\cite{Mann:1998}. However, these results include only the contribution of a single massless scalar field. When 1-loop corrections from all massless fields are summed together, the entropy of a non-extremal Reissner-Nordstr\"{o}m black hole obeys~\cite{Sen:2013}
\begin{align}
\label{full entanglement entropy}
 S= & \frac{k_B \mathcal{A}}{4l_{P}^2}+k_B\mathcal{C}\ln\left(\frac{\mathcal{A}}{\mathcal{A}_0}\right),
\end{align}
where $\mathcal{C}$ is an explicitly known function that depends on the black hole's mass, $M$, and charge, $Q$, as well as on the number of fields and their spins. In the case of the standard model of particle physics, the logarithmic term is positive for any $Q$. However, the equation breaks down for an extremal black hole ($Q=M$ in geometrized units). Therefore, one cannot use it for the case of a 2-sphere in flat spacetime (topologically equivalent to an extremal Reissner-Nordstr\"{o}m black hole), which would have to be treated separately.

Regarding other approaches to entropy calculations, LQG and AdS/CFT yield a negative sign of the logarithmic term~\cite{Meissner:2004,Carlip:2000}. In string theory, the sign varies for different variants of the theory~\cite{Sen:2013}. In~\cite{Hod:2004}, assumptions about quantisation of the horizon area are used to argue that the coefficient in front of the logarithm is a positive integer. The analysis of canonical corrections in~\cite{Medved:2005} also yields a positive sign.

Generally speaking, corrections to the Bekenstein formula due to quantum gravity effects can be divided into two categories. Microcanonical corrections are found by more precise counting of the microstates at fixed horizon area. Naturally, such a procedure reduces our uncertainty and, therefore, leads to negative corrections to the entropy~\cite{Medved:2004,Medved:2005,Alonso:2018}. The second category, canonical corrections, arise due to thermal fluctuations of the horizon area at a fixed temperature. Since fluctuations are an additional source of uncertainty, it follows that canonical corrections to entropy have a positive sign~\cite{Medved:2004,Medved:2005,Alonso:2018}. Due to this sign ambiguity we will consider the most general form of the quantum modified Bekenstein entropy~\eqref{modified_BH} in the following.

\section{Modified equations of motion}
\label{modified_equations}

In this section, we employ tools developed in thermodynamics of spacetime together with the logarithmic correction to entanglement entropy in order to obtain modifications to the gravitational equations of motion. Since the logarithmic term arises as a leading quantum gravity effect (see subsection~\ref{entanglement_log}), we, in fact, explore the effective dynamics at the low energy limit of quantum gravity. Nevertheless, we still assume spacetime to be describable as a 4-dimensional Lorentzian manifold, i.e. we limit ourselves to length scales significantly larger than the Planck length. This allows us to directly apply the already known methods of semiclassical thermodynamics of spacetime.

While derivation of Einstein equations seems to work equally well for Rindler horizons, causal diamonds and light cones (and probably other types of local causal horizons as well), logarithmic corrections to entanglement entropy only appear for closed causal horizons as we have previously discussed. Since the spherical symmetry of the horizon immensely simplifies the calculations, we essentially have only two kinds of objects worth considering, a light cone and a causal diamond. The light cones were used to derive Einstein equations as well as equations of motion for some modified theories of gravity~\cite{Svesko:2017}. We instead choose causal diamonds, as they form a closed region of spacetime with a naturally defined boundary, making them very well suited for local calculations. However, we expect that the results for null cones and causal diamonds should be equivalent, as the spatial cross-section of the causal horizon is a 2-sphere in both cases and both yield the same results in the semiclassical case. Including the logarithmic corrections in light cone thermodynamics might thus be a useful double-check of our results.

We carry out the derivation by two independent methods. The first one generalises the derivation of Einstein equations from the maximal vacuum entanglement hypothesis~\cite{Jacobson:2015}. The second is based on their derivation from the Clausius entropy flux~\cite{Alonso-Serrano-Liska}. However, to develop these derivations, we first need to introduce a basic geometric tool, the geodesic local causal diamond.

\subsection{Geodesic local causal diamonds}
\label{GLCD geometry}

The thermodynamic objects that we will use to obtain gravitational dynamics are geodesics local causal diamonds (GLCD). Here, we briefly present their construction and basic geometric properties. More detailed treatment of the causal diamonds is provided, e.g. \mbox{in~\cite{Gibbons:2007,Jacobson:2017,Wang:2019}.}
	
\begin{figure}[tbp]
\centering
\includegraphics[width=.45\textwidth,origin=c,trim={0.1cm 2.4cm 36.7cm 1.5cm},clip]{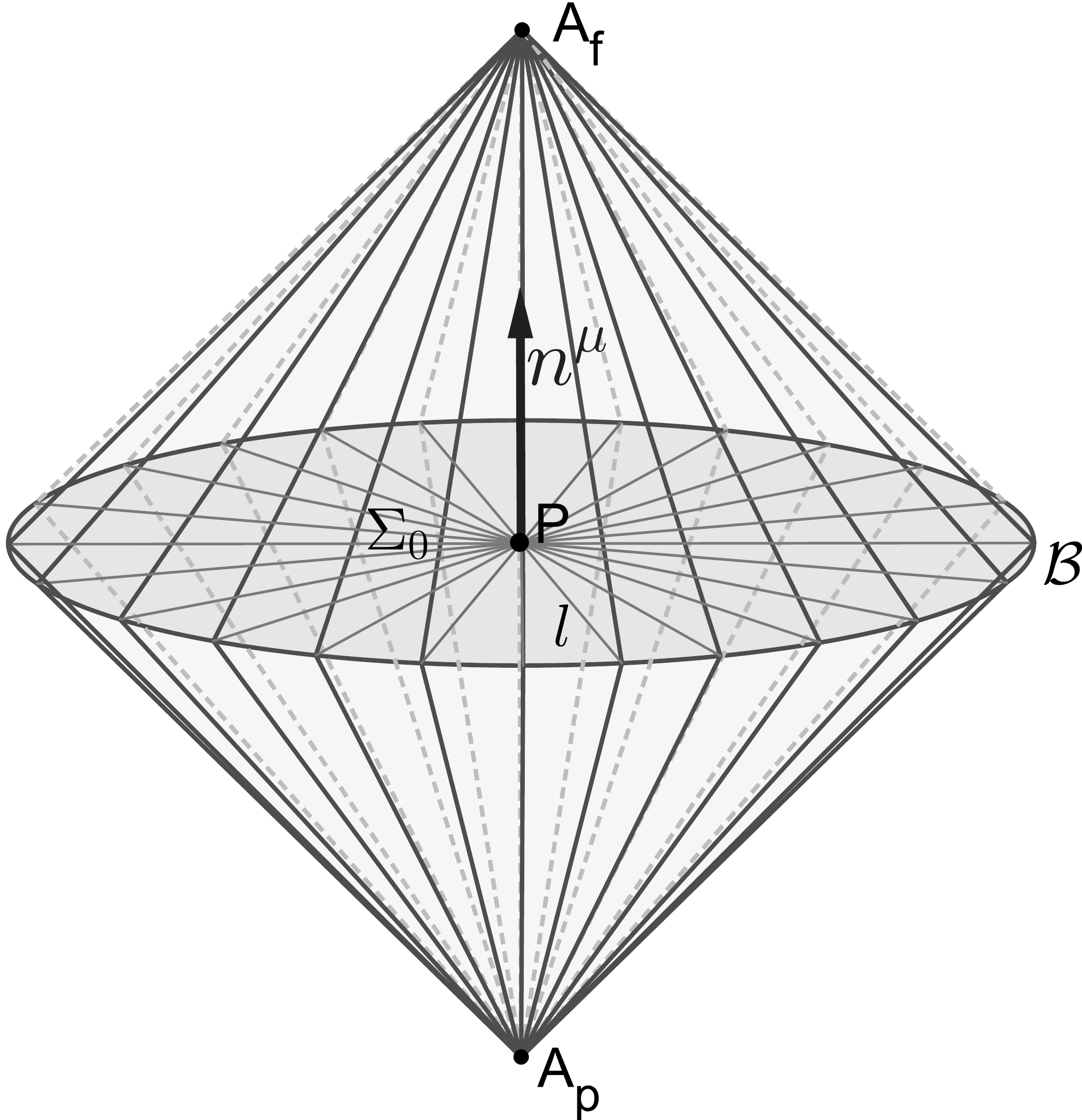}
\caption{\label{diamond} A sketch of a GLCD with the origin in point $P$ (the angular coordinate $\theta$ is suppressed). $\Sigma_0$ is a spatial geodesic ball of radius $l$ (several of the geodesics forming it are depicted as grey lines), its boundary an approximate 2-sphere $\mathcal{B}$. Unit timelike vector $n^{\mu}$ is a normal of $\Sigma_0$. The tilted lines from the past apex $A_p$ ($t=-l/c$) to the future apex $A_f$ ($t=l/c$) represent the null geodesic generators of the GLCD boundary. The diamond's base $\Sigma_0$ is the spatial cross-section of both the future domain of dependence of $A_p$ and the past domain of dependence of $A_f$ at $t=0$.}
\end{figure}

In an arbitrary point $P$ of spacetime pick any unit timelike vector $n^{\mu}$. Choose Riemann normal coordinates (RNC) so that it holds $n=\partial/\partial t+O\left(l\right)$. The RNC metric expansion around $P$ yields~\cite{Brewin:2009}
\begin{equation}
\label{RNC}
g_{\mu\nu}(x)=\eta_{\mu\nu}-\frac{1}{3}R_{\mu\alpha\nu\beta}\left(P\right)x^{\alpha}x^{\beta}+O\left(x^3\right).
\end{equation}
The family  of geodesics orthogonal to $n^{\mu}$ departing from $P$ with parameter length $l$, form an approximate 3-dimensional ball, $\Sigma_0$. This ball causally determines a region of spacetime called a geodesic local causal diamond (see figure~\ref{diamond}). For a small enough $l$ (much smaller than the local curvature length), the boundary $\mathcal{B}$ of $\Sigma_0$ is an approximate 2-sphere of area~\cite{Jacobson:2015}
\begin{equation}
\label{area}
\mathcal{A}=4\pi l^2-\frac{4\pi}{9}l^4G_{00}\left(P\right)+O\left(l^5\right),
\end{equation}
where $G_{00}\equiv G_{\mu\nu}n^{\mu}n^{\nu}$.
	
It is straightforward to show that a GLCD possesses (up to $O\left(l^3\right)$ terms) a spherically symmetric conformal isometry generated by a conformal Killing vector~\cite{Jacobson:2015} given by
\begin{equation}
\label{conformal Killing}
\xi=C\left(\left(l^2-t^2-r^2\right)\frac{\partial}{\partial t}-2rt\frac{\partial}{\partial r}\right),
\end{equation}
where $C$ denotes an arbitrary normalisation constant. The null boundary of a GLCD thus corresponds to a conformal Killing horizon. As we will see in the following subsection, the presence of a conformal Killing vector allows us to define the variation of matter entanglement entropy inside $\Sigma_0$.

\subsection{Derivation from MVEH}
\label{MVEH_modified_equations}

Einstein equations can be derived by considering a variation of entanglement entropy of a GLCD around equilibrium~\cite{Jacobson:2015}. The main ingredient of the derivation is the maximal vacuum entanglement hypothesis (MVEH): ``When the geometry and quantum fields are simultaneously varied from maximal symmetry, the entanglement entropy in a small geodesic ball is maximal at fixed volume''~\cite{Jacobson:2015}. The formulation of the MVEH implicitly demands a finite and universal prescription for the area density of the vacuum entanglement entropy. If the prescription for $S_e$ coincides with the Bekenstein formula, the MVEH applied to a first order variation of the quantum fields from their vacuum state implies Einstein equations~\cite{Jacobson:2015}. To obtain quantum modifications to gravitational dynamics, we instead consider the modified entropy with a logarithmic term. The condition of maximal entanglement entropy can be mathematically stated as $\delta S_{e,q}+\delta S_m=0$, where $S_{e,q}$ denotes modified vacuum entanglement entropy, $S_m$ entanglement entropy of the matter and we consider a first order variation of both metric and quantum fields from vacuum, maximally symmetric spacetime. We evaluate $\delta S_{e,q}$ using the expression for the area of $\mathcal{B}$ found in the previous subsection and express $\delta S_m$ in terms of variation of the energy-momentum tensor expectation value, $\delta\langle T_{\mu\nu}\rangle$. To carry out this derivation, we need to consider possible modifications of Unruh temperature, conformality (or not) of the fields, applicability of Einstein equivalence principle (EEP) and universality of $G$ (that automatically holds if strong equivalence principle (SEP) holds). Then, the maximal entanglement entropy condition implies (albeit with some technical caveats) modified gravitational equations of motion
\begin{equation}
\label{quantum EoM entanglement}
S_{\mu\nu}-\frac{Cl_{Pl}^2}{30\pi}S_{\mu\lambda}S^{\lambda}_{\;\:\nu}+\frac{Cl_{Pl}^2}{120\pi}\left(R_{\kappa\lambda}R^{\kappa\lambda}-\frac{1}{4}R^2\right)g_{\mu\nu}=\frac{8\pi G}{c^4}\left(\delta\langle T_{\mu\nu}\rangle-\frac{1}{4}\delta\langle T\rangle g_{\mu\nu}\right),
\end{equation}
where $S_{\mu\nu}=R_{\mu\nu}-Rg_{\mu\nu}/4$ denotes the traceless part of Ricci tensor. It is worth noting that these equations present a direct generalisation of the equations of motion of UG rather than GR, just as it was found in the semiclassical case~\cite{Alonso-Serrano-Liska}. The curvature terms should be understood as quantum expectation values, since the energy-momentum tensor is also interpreted in this way~\cite{Jacobson:2015,Alonso-Serrano-Liska}. In the following and until the end of the subsection, we discuss the derivation of these equations in detail.

We start by specifying the appropriate equilibrium state of geometry. Even in the semiclassical setting, it turns out to be a maximally symmetric spacetime with $G^{MSS}_{\mu\nu}=-\lambda g_{\mu\nu}$, rather than flat spacetime~\cite{Jacobson:2015,Alonso-Serrano-Liska}. The curvature scale $\lambda$ is approximately constant inside the GLCD, but in general depends on both its origin $P$ and size parameter $l$~\cite{Jacobson:2015}. Based on the semiclassical case~\cite{Alonso-Serrano-Liska}, we expect that $\lambda$ consists of three separate contributions
\begin{equation}
\lambda=\lambda_0+\lambda_{nc}+\lambda_{q},
\end{equation}
where $\lambda_0=-R/4$ is the semiclassical value of $\lambda$ when only conformal fields are considered, $\lambda_{nc}$ contains terms coming from the presence of non-conformal fields\footnote{Semiclassically, it holds $\lambda_{nc}=\left(8\pi G/c^4\right)\left(\delta\langle T\rangle/4-\delta X\right)$~\cite{Alonso-Serrano-Liska}.} and $\lambda_{q}$ arises due to quantum corrections and is proportional to $l_P^2$.

Consider a GLCD constructed in a maximally symmetric spacetime (MSS). A variation of the quantum fields from their vacuum state induces a variation of the metric and, consequently, of entanglement entropy. The modified entanglement entropy $S_{e,q}$ associated with the GLCD horizon equals
\begin{equation}
\label{S_UV_log}
S_{e,q}=\eta\mathcal{A}+k_{B}\mathcal{C}\ln \frac{\mathcal{A}}{\mathcal{A}_0} + O\left(\frac{k_B l_P^2}{\mathcal{A}}\right),
\end{equation}
where $\mathcal{C}$ is a dimensionless constant. To be as general as possible, we assume an undetermined proportionality constant $\eta$ in the leading term instead of fixing it to the Bekenstein value $k_B/4l_P^2$. The variation of the metric changes the GLCD horizon area and entanglement entropy $S_{e,q}$ associated with it. One has to consider a variation of the metric that leaves fixed the volume of $\Sigma_0$~\cite{Jacobson:2015,Bueno:2017}. Otherwise, Einstein equations cannot be recovered in the limit $\mathcal{C}\to0$. The fixed volume variation of area obeys
\begin{equation}
\label{delta A}
\delta\mathcal{A}\vert_{V}=-\frac{4\pi}{15}l^4\left(G_{00}\left(P\right)+\lambda\left(P\right)g_{00}\left(P\right)\right)+O\left(l^5\right).
\end{equation}
The corresponding change of $S_{e,q}$ equals
\begin{align}
\nonumber \delta S_{e,q} &=S_{e,q}-S^{MSS}_{e,q}\\
&=\eta\delta\mathcal{A}+k_{B}\mathcal{C}\frac{\delta\mathcal{A}}{\mathcal{A}_{MSS}}-k_B\frac{\mathcal{C}}{2}\left(\frac{\delta\mathcal{A}}{\mathcal{A}_{MSS}}\right)^2+O\left(\left(\delta\mathcal{A}\right)^3\right).
\end{align}

Next, we turn our attention to the variation of entropy of the quantum fields from vacuum. A small variation from the vacuum state leads to a change of the matter entanglement entropy, $\delta S_m$, in the spatial geodesic ball $\Sigma_0$. The vacuum state of the field can be expressed in terms of a thermal density matrix at the Unruh temperature, from which one obtains the entropy variation. In the semiclassical set-up (without any quantum gravity corrections), $\delta S_m$ is given by (for a detailed derivation see~\cite{Jacobson:2019})
\begin{equation}
\label{non-conformal matter entropy}
\delta S_{m}=\frac{2\pi k_B}{\hbar c}\frac{4\pi l^4}{15}\left(\delta\langle T_{00}\rangle+\delta X\right),
\end{equation}
where $\delta X$ indicates the variation of a spacetime scalar that in general depends on $l$ and is present only when non-conformal fields are considered~\cite{Jacobson:2019}. This equation holds only for fields with a fixed UV point.

Since $\delta S_m$ directly depends on the Unruh temperature, its possible modifications due to quantum gravity effects need to be considered. Recently, quantum gravity corrections to the Unruh temperature formula were suggested in the context of GUP~\cite{Scardigli:2018jlm,Luciano:2019}, following earlier proposal of such corrections to Hawking temperature~\cite{Adler:2001}. While several slightly different formulas appear in the literature, all include a leading order correction term proportional to $T^3\propto a^3$, with $T$ being the standard Unruh temperature. Thus, all of them can be written in the following form
\begin{equation}
\label{modified_Unruh}
T_{GUP}=\frac{\hbar a\left(1+\psi\frac{l_P^2}{c^4}a^2\right)}{2\pi k_Bc}+O\left(\frac{l_P^4a^5}{c^8}\right),
\end{equation}
where $\psi$ is a real number that can be expected to be of the order of unity~\cite{Scardigli:2018jlm,Luciano:2019}. The existence of any such corrections to the Unruh (or even Hawking) temperature remains uncertain and, to our best knowledge, the modified temperature was not reproduced by any rigorous method. However, given the rather general grounds on which GUP is expected to hold~\cite{Garay:1994en}, corrections of this form are worth considering. At the very least, we should prove that our construction does not break down if they appear. Therefore, we work with the modified temperature in the following. By setting $\psi=0$, we recover the standard Unruh temperature, which is thus considered as a special case. In fact, we demonstrate that gravitational dynamics are independent of the exact form of the Unruh temperature (assuming it is polynomial in acceleration $a$). As both Hawking and Unruh effects are purely kinematic phenomena unrelated to gravitational equations of motion~\cite{Visser:2003}, this is expected.

To consider the standard Unruh effect, one needs to assume that the ground state of quantum fields is locally approximated by Minkowski vacuum. This amounts to invoking Einstein equivalence principle~\cite{Chirco:2010}: ``Fundamental non-gravitational test physics is not affected, locally and at any point of spacetime, by the presence of a gravitational field~\cite{Casola:2015}.'' However, when quantum gravity effects become relevant, the status of EEP is unclear. For example, possible violations of weak equivalence principle (WEP, a necessary condition for EEP) due to GUP phenomenology were investigated by numerous works~\cite{Ghosh:2013,Tkachuk:2013,Pramanik:2014,Casadio:2020} with very different conclusions. However, by considering a modified Unruh temperature, we account for changes in the local quantum field theory induced by a possible EEP violation. This allows us to consider the (modified) Unruh temperature regardless of the validity of EEP. We will return to the question of equivalence principles in subsection~\ref{Invariant?}.

The variation of $S_{m}$ then satisfies
\begin{equation}
\label{modified S_m}
\delta S_{m}=\frac{2\pi k_Bc}{\hbar}\frac{4\pi}{15}l^4\left(\delta\langle T_{00}\rangle+\delta X\right)-4\psi l^2C^2\frac{2\pi k_Bl_P^2}{\hbar c^3}\frac{4\pi}{15}l^4\left(\delta\langle T_{00}\rangle+\delta X\right)+O\left(l^5\right),
\end{equation}
where $C$ is an arbitrary normalisation of the conformal Killing vector. When the standard Unruh temperature is considered, $C$ does not affect the matter entanglement entropy and it is possible to fix it to any value (usually, $C=1/2l$ is chosen, corresponding to the unit surface gravity~\cite{Jacobson:2015,Jacobson:2019}). However, for $\psi\ne0$, the modified temperature explicitly depends on $C$. As we cannot motivate choosing any value of $C$, we keep it as an arbitrary constant.

Upon computing both variations, we invoke the MVEH, which demands that the total entropy variation vanishes to the first order, i.e. $\delta S_{e,q}+\delta S_m=0$,
\begin{align}
\nonumber & -\eta\left(S_{00}\left(P\right)-\lambda_{nc}-\lambda_q\right)-\frac{k_B\mathcal{C}}{120\pi}\left(S_{00}\left(P\right)-\lambda_{nc}-\lambda_{q}\right)^2-\frac{k_B\mathcal{C}}{4\pi l^2}\left(S_{00}\left(P\right)-\lambda_{nc}-\lambda_{q}\right) \\
 & +\frac{2\pi k_Bc}{\hbar}\left(1-4\psi\frac{l_P^2}{c^4}l^2C^2\right)\left(\delta\langle T_{00}\rangle+\delta X\right)+O\left(l\right)=0, \label{balance_GUP_1}
\end{align}
This equation holds for any $l$ much smaller than all the relevant length scales of the quantum field theory, the excitation length and the local curvature length (i.e., inverse of the square root of the Riemann's tensor largest eigenvalue), but still much larger than Planck length. In the range allowed by these constraints, $l$ can be arbitrary. Thus, the MVEH condition depends continuously on $l$, and it can be differentiated with respect to it, yielding
\begin{align}
\nonumber &-\frac{\text{d}}{\text{d}l}\bigg[\eta\left(\lambda_{nc}+\lambda_q\right)+\frac{k_B\mathcal{C}}{60\pi}\left(S_{00}\left(P\right)-\lambda_{nc}-\lambda_{q}\right)\left(\lambda_{nc}+\lambda_{q}\right)-\frac{k_B\mathcal{C}}{4\pi l^2} &\\
& \left(S_{00}\left(P\right)-\lambda_{nc}-\lambda_{q}\right)+\frac{2\pi k_Bc}{\hbar}\left(1-4\psi\frac{l_P^2}{c^4}l^2C^2\right)\left(\delta\langle T_{00}\rangle+\delta X\right)+O\left(l\right)\bigg] =0 \label{d/dl},&
\end{align}
as metric, curvature and energy-momentum tensors are evaluated in $P$ and their values cannot depend on $l$. To satisfy this condition, it must hold
\begin{eqnarray}
\nonumber &\eta\left(\lambda_{nc}+\lambda_q\right)+\frac{k_B\mathcal{C}}{60\pi}\left(S_{00}\left(P\right)-\lambda_{nc}-\lambda_{q}\right)\left(\lambda_{nc}+\lambda_{q}\right)-\frac{k_B\mathcal{C}}{4\pi l^2}\big(S_{00}\left(P\right) &\\
& -\lambda_{nc}-\lambda_{q}\big)+\frac{2\pi k_Bc}{\hbar}\left(1-4\psi\frac{l_P^2}{c^4}l^2C^2\right)\left(\delta\langle T_{00}\rangle+\delta X\right)+O\left(l\right)=-\Phi\left(P\right).& \label{Phi}
\end{eqnarray}
Here $\Phi$ is a scalar independent of $l$.

To be consistent with semiclassical results~\cite{Jacobson:2019,Alonso-Serrano-Liska}, we need to recover Einstein equations in the limit of $\mathcal{C}\to0$. Clearly, this requirement fixes $\eta=k_B/4l_P^2$. In order to regard $\eta$ as a universal constant, one needs to consider that $G$ has a constant value throughout the spacetime. This can be done by invoking the strong equivalence principle~\cite{Chirco:2010,Alonso-Serrano-Liska}: ``All test fundamental physics (including gravitational physics) is not affected, locally, by the presence of a gravitational field''~\cite{Casola:2015}. However, given the above discussed problems with equivalence principles when quantum gravity effects are considered, we limit ourselves to demand that $G$ is a universal constant, instead of assuming SEP. We will further address this issue in section~\ref{interpretation}.

Since the time indices are, in fact, contractions with timelike vector $n^{\mu}$, we have
\begin{equation}
\label{hc}
S_{\mu\nu}\left(P\right)n^{\mu}n^{\nu}+\frac{\mathcal{C}l_{Pl}^2}{30\pi}S_{\alpha\beta}\left(P\right)n^{\alpha}n^{\beta}S_{\mu\nu}\left(P\right)n^{\mu}n^{\nu}-\Phi\left(P\right)=\frac{8\pi G}{c^4}\delta\langle T_{\mu\nu}\left(P\right)\rangle\left(P\right)n^{\mu}n^{\nu}.
\end{equation}
This equation holds for any unit timelike vector $n^{\mu}$ defined in $P$ and there is no preferred time direction in this construction. As the dependence on $n^{\mu}$ is continuous, we can differentiate it with respect to it. This yields a system of conditions (for simplicity of notation, we do not write the dependences on $P$ in the following)
\begin{align}
\frac{\partial^5\Phi}{\partial n^{\kappa}\partial n^{\lambda}\partial n^{\mu}\partial n^{\nu}\partial n^{\rho}} &=0, \label{c1}\\
\frac{\partial^4\Phi}{\partial n^{\kappa}\partial n^{\lambda}\partial n^{\mu}\partial n^{\nu}} &=24\frac{\mathcal{C}l_{P}^2}{30\pi}S_{(\kappa\lambda}S_{\mu\nu)},  \label{c2}\\
\frac{\partial^3\Phi}{\partial n^{\lambda}\partial n^{\mu}\partial n^{\nu}} &=24\frac{\mathcal{C}l_{P}^2}{30\pi}S_{(\kappa\lambda}S_{\mu\nu)}n^{\kappa},\\
\frac{\partial^2\Phi}{\partial n^{\mu}\partial n^{\nu}} &=2S_{\mu\nu}+12\frac{\mathcal{C}l_{P}^2}{30\pi}S_{(\kappa\lambda}S_{\mu\nu)}n^{\kappa}n^{\lambda}-2\frac{8\pi G}{c^4}\delta\langle T_{\mu\nu}\rangle,  \label{c4}\\
\frac{\partial\Phi}{\partial n^{\mu}} &=2S_{\mu\nu}n^{\nu}+4\frac{\mathcal{C}l_{P}^2}{30\pi}S_{(\kappa\lambda}G_{\mu\nu)}n^{\kappa}n^{\lambda}n^{\nu}-2\frac{8\pi G}{c^4}\delta\langle T_{\mu\nu}\rangle n^{\nu}.
\end{align}
The first condition guarantees that the second one is already completely independent of $n^{\mu}$. It can be solved either by $\left(\mathcal{C}l_{Pl}^2/30\pi\right)S_{(\kappa\lambda}S_{\mu\nu)}n^{\mu}n^{\nu}h^{\kappa\lambda}$, or $\left(\mathcal{C}l_{Pl}^2/30\pi\right)S_{(\kappa\lambda}S_{\mu\nu)}n^{\mu}n^{\nu}n^{\kappa}n^{\lambda}$, with $h^{\kappa\lambda}=g^{\kappa\lambda}+n^{\kappa}n^{\lambda}$ being a three metric on hypersurface orthogonal to $n^{\mu}$. No other solution that can be constructed from the matter and metric variables having correct dimensions (note that the only constant with the dimension of length is $l_{Pl}$ and any $O\left(l_{Pl}^3\right)$ terms are already neglected). Choosing $\left(\mathcal{C}l_{Pl}^2/30\pi\right)S_{(\kappa\lambda}S_{\mu\nu)}n^{\mu}n^{\nu}n^{\kappa}n^{\lambda}$, we would just reproduce Einstein equations. However, we have seen that our procedure clearly yields a non-trivial correction term. Furthermore, at least in the terms proportional to $l^4$, only tensors projected onto $n^{\mu}$ in all indices influence the change of the diamond's area and, thus, its entropy~\footnote{This was confirmed even for the $O\left(l^6\right)$ terms in the vacuum case by the analysis of the higher-order corrections to the area of a GLCD~\cite{Jacobson:2017}. The result in 4 dimensions is proportional to $t_{\kappa\lambda\mu\nu}n^{\kappa}n^{\lambda}n^{\mu}n^{\nu}$, where $t_{\kappa\lambda\mu\nu}=C_{\kappa\alpha\mu\beta}C_{\lambda\;\:\nu}^{\;\:\alpha\;\:\beta}+6*C_{\kappa\alpha\mu\beta}*C_{\lambda\;\:\nu}^{\;\:\alpha\;\:\beta}$, with $C_{\kappa\alpha\mu\beta}$ being the Weyl tensor and $*$ denoting the Hodge dual.}. Therefore, any projections to the hypersurface orthogonal to $n^{\mu}$ will not contribute to our entropy balance equation. Previous arguments allow us to set
\begin{equation}
\Phi=\frac{\mathcal{C}l_{Pl}^2}{30\pi}S_{\kappa\mu}S_{\lambda\nu}n^{\mu}n^{\nu}h^{\kappa\lambda}+{^{(3)}\Phi}_{\lambda\mu\nu}n^{\lambda}n^{\mu}n^{\nu}+{^{(2)}\Phi}_{\mu\nu}n^{\mu}n^{\nu}+{^{(1)}\Phi}_{\mu}n^{\mu}+{^{(0)}\Phi}, 
\end{equation}
where ${^{(3)}\Phi}_{\lambda\mu\nu}$, ${^{(2)}\Phi}_{\mu\nu}$, ${^{(1)}\Phi}_{\mu}$ and ${^{(0)}\Phi}$ are tensors independent of $n^{\mu}$. However, no tensors of the form ${^{(3)}\Phi}_{\lambda\mu\nu}$, ${^{(1)}\Phi}_{\mu}$ can be constructed just from the metric, the curvature tensors and the energy-momentum tensor. Condition~\eqref{c4} in principle allows terms in ${^{(2)}\Phi}_{\mu\nu}$ proportional to $l_{P}^2RR_{\mu\nu}$, $l_{P}^2R_{\mu}^{\;\:\lambda}R_{\lambda\nu}$, $\left(l_{P}^2G^2/c^8\right)\delta\langle T_{\mu}^{\;\:\lambda}\rangle\delta\langle T_{\lambda\nu}\rangle$, $\left(l_{P}^2G^2/c^8\right)\delta\langle T\rangle\delta\langle T_{\mu\nu}\rangle$, $\left(l_{P}^2G/c^4\right)R\delta\langle T_{\mu\nu}\rangle$, $\left(l_{P}^2G^2/c^8\right)\delta\langle T_{\mu}^{\;\:\lambda}\rangle R_{\lambda\nu}$, $\left(l_{P}^2G/c^4\right)\delta\langle T\rangle R_{\mu\nu}$ and various terms involving the Weyl tensor. However, any such terms would appear in our entropy balance equation, projected on $n^{\mu}$ in both indices (in other words, they contain no projections to the hypersurface orthogonal to $n^{\mu}$, that can be missed by our procedure). The previous arguments reduce the unknown terms in the equilibrium condition to a single undetermined scalar function ${^{(0)}\Phi}$
\begin{equation}
\label{hamiltonian constraint}
\left(S_{\mu\nu}\left(P\right)-\frac{Cl_{Pl}^2}{30\pi}S_{\mu\lambda}\left(P\right)S^{\lambda}_{\;\:\nu}\left(P\right)-\frac{8\pi G}{c^4}T_{\mu\nu}\left(P\right)\right)n^{\mu}n^{\nu}={^{(0)}\Phi}\left(P\right),
\end{equation}
which holds for any unit timelike vector $n^{\mu}$ defined in point $P$. Following an argument similar to the one the authors explained in detail in~\cite{Alonso-Serrano-Liska}, we can dispense with contractions with $n^{\mu}$ and the dependence on $P$ (invoking EEP), and obtain a system of equations valid throughout spacetime
\begin{equation}
\label{eom}
S_{\mu\nu}-\frac{Cl_{Pl}^2}{30\pi}S_{\mu\lambda}S^{\lambda}_{\;\:\nu}-\frac{8\pi G}{c^4}\delta\langle T_{\mu\nu}\rangle=-{^{(0)}\Phi}g_{\mu\nu}.
\end{equation}
Lastly, we determine ${^{(0)}\Phi}$ by taking a trace of the equations. Then, we obtain traceless equations of gravitational dynamics\footnote{Let us note that, while the procedure used to obtain the equations of motion from equilibrium condition does lead to some ambiguities in the correction terms, they all disappear for $R=T=0$, i.e., for situations in which matter sources are only conformal fields and the cosmological constant is zero (or small enough not to affect the correction terms). Then, one unambiguously finds
	\begin{equation}
	R_{\mu\nu}-\frac{Cl_{Pl}^2}{30\pi}R_{\mu\lambda}R^{\lambda}_{\;\:\nu}+\frac{Cl_{Pl}^2}{120\pi}R_{\kappa\lambda}R^{\kappa\lambda}g_{\mu\nu}=\frac{8\pi G}{c^4}\delta\langle T_{\mu\nu}\rangle.
	\end{equation}}
\begin{equation}
\nonumber S_{\mu\nu}-\frac{Cl_{Pl}^2}{30\pi}S_{\mu\lambda}S^{\lambda}_{\;\:\nu}+\frac{Cl_{Pl}^2}{120\pi}\left(R_{\kappa\lambda}R^{\kappa\lambda}-\frac{1}{4}R^2\right)g_{\mu\nu}=\frac{8\pi G}{c^4}\left(\delta\langle T_{\mu\nu}\rangle-\frac{1}{4}\delta\langle T\rangle g_{\mu\nu}\right).
\end{equation}
Before further discussing implications of modified equations for gravitational dynamics, we will present an alternative derivation from the Clausius entropy flux based on the semiclassical derivation previously developed by the authors~\cite{Alonso-Serrano-Liska}.

\subsection{Derivation from the Clausius entropy flux}
\label{clausius_modified_equations}

The second method of derivation of effective equations of motion we consider, uses equilibrium condition on the flux of energy-momentum across the null boundary of a GLCD. It can be stated as $\Delta S_{e,q}+\Delta S_C=0$, where $S_{e,q}$ is the same as in previous subsection and $S_C$ denotes the flux of Clausius entropy. For a GLCD, $\Delta S_C$ turns out to be proportional to the traceless part of energy-momentum tensor~\cite{Baccetti:2013ica,Alonso-Serrano-Liska}. After some calculations, modifying the definition of Clausius entropy to account for a possible modified Unruh temperature in addition to considerations in previous derivation, the equilibrium condition then leads to modified gravitational equations of motion 
\begin{equation}
\label{quantum EoM clausius}
S_{\mu\nu}-\frac{\mathcal{C}l_{P}^2}{18\pi}S_{\mu\lambda}S^{\lambda}_{\;\:\nu}+\frac{\mathcal{C}l_{P}^2}{72\pi}\left(R_{\kappa\lambda}R^{\kappa\lambda}-\frac{1}{4}R^2\right)g_{\mu\nu}=\frac{8\pi G}{c^4}\left(T_{\mu\nu}-\frac{1}{4}Tg_{\mu\nu}\right),
\end{equation}
that are nearly identical to the ones obtained from the MVEH, except for a different proportionality constant in front of the correction terms ($-\mathcal{C}/18\pi$ instead of $-\mathcal{C}/30\pi$). We now again provide a detailed discussion of this derivation for the rest of the subsection.

A general prescription for the observer-dependent Clausius entropy of a bifurcate null surface states~\cite{Baccetti:2013ica}
\begin{equation}
\label{clausius_arbitrary}
S_{C}\left(\lambda\right)=S\left(\mathcal{B}\right)+\frac{2\pi k_Bc}{\hbar}\int_{0}^{\lambda}\int_{\mathcal{S}\left(\lambda\right)}\tilde{\lambda}T_{\mu\nu}k_{\pm}^{\mu}k_{\pm}^{\nu}\text{d}^2\mathcal{A}\text{d}\tilde{\lambda}+O\left(\lambda^3\right),
\end{equation}
where $\lambda$ is the affine parameter along the geodesic generators of the null surface (in general, it can be different for each generator), $k_{\pm}^{\mu}$ are null vectors tangent to the surface for $\lambda>0$ and $\lambda<0$, respectively, $\text{d}^2\mathcal{A}$ denotes the area element of the null surface's spatial cross-section $\mathcal{S}\left(\lambda\right)$, and $S\left(\mathcal{B}\right)$ is an undetermined Clausius entropy associated with the spatial bifurcation 2-surface $\mathcal{B}$ corresponding to $\lambda=0$.

To obtain quantum modified gravitational dynamics from the Clausius entropy flux, we must first show that some prescription for Clausius entropy exist even when the Unruh temperature is modified in the way discussed in the previous subsection. The standard semiclassical construction of Clausius entropy~\cite{Baccetti:2013ica} no longer works. In the following, we modify it to make it still applicable. While we concentrate on the specific case of GLCD's, a generalisation to an arbitrary bifurcate null surface can be done along the same lines as in the semiclassical case.

Consider a class of timelike observers travelling inside GLCD with constant acceleration~$a$. Coordinates on their worldlines read
\begin{equation}
x^{\mu}(\tau,\theta,\phi)=\left(\frac{c^2}{a}\sinh{\left(\frac{a\tau}{c^2}\right)},l-\frac{c^2}{a}\cosh{\left(\frac{a\tau}{c^2}\right)},\theta,\phi\right)+O\left(l^2\right),
\end{equation}
where $\tau$ is the proper time they measure, $\theta$, $\phi$ are the usual angular coordinates and $O\left(l^2\right)$ corrections come from the higher-order terms in the RNC expansion of the metric (we use that, inside GLCD, $\tau\le l$ in writing the error terms). Such observers have the following velocity and acceleration
\begin{align}
V^{\mu}=c\left(\cosh{\left(\frac{a\tau}{c^2}\right)},-\sinh{\left(\frac{a\tau}{c^2}\right)},0,0\right)+O\left(l\right), \label{velocity}\\
a^{\mu}=a\left(\sinh{\left(\frac{a\tau}{c^2}\right)},-\cosh{\left(\frac{a\tau}{c^2}\right)},0,0\right)+O\left(1\right).
\end{align}
The normal of the timelike hyperbolic sheet $\Sigma$ these observers sweep out is given by
\begin{equation}
\label{normal}
N^{\mu}=\left(-\sinh{\left(\frac{a\tau}{c^2}\right)},\cosh{\left(\frac{a\tau}{c^2}\right)},0,0\right)+O\left(l\right).
\end{equation}
Using the above defined vectors, heat (matter-energy) crossing a segment of $\Sigma$ can be written in terms of the energy-momentum tensor~\cite{Baccetti:2013ica}
\begin{equation}
\label{heat}
\delta Q=-\frac{1}{c}\int_{\Sigma}T_{\mu\nu}V^{\mu}N^{\nu}\text{d}^3\Sigma.
\end{equation}
Since the (modified) Unruh temperature is measured by uniformly accelerating observers, we can define entropy of the energy crossing it by the equilibrium Clausius relation, $\text{d}S_{C}=\delta Q/T_{GUP}$.

Originally, entropy is computed in the limit of $a\to\infty$~\cite{Baccetti:2013ica}. Then, $\Sigma$ approaches the bifurcate null surface. However, when one considers the modified Unruh temperature, the term proportional to $a^2$ becomes dominant as $a$ approaches infinity. To prevent this, we allow $a$ to become very large but still smaller than $1/\sqrt{\psi}$. Then, $\Sigma$ approaches the causal horizon close enough to ignore the difference between them, but the correction term in the modified Unruh temperature remains sub-leading.

The time derivative of heat crossing $\Sigma$ in the limit of large $a$ equals
\begin{equation}
\label{dQ/dt}
\frac{\delta Q\left(t\right)}{\text{d}t}=\int_{\mathcal{S}\left(t\right)}\lbrace T_{tt}+T_{rr}-2\,\textrm{sign}(t)T_{tr}\rbrace at\text{d}^2\mathcal{A}+O\left(l^4\right)+O\left(\frac{1}{a}\right),
\end{equation}
where ${\mathcal{S}\left(t\right)}$ is a spatial cross-section of $\Sigma$ with a constant value of the time coordinate, $t$, and $\text{d}^2\mathcal{A}\left(t\right)=\left(l-t\right)^2\text{d}^2\Omega+O\left(l^4\right)$ is the area element on ${\mathcal{S}\left(t\right)}$. This can be written as an integral of $T_{\mu\nu}k_{\pm}^{\mu}k_{\pm}^{\nu}$, where $k_{\pm}^{\mu}$ are null vectors corresponding to the positive and negative values of $t$, respectively. They satisfy
\begin{equation}
\label{null}
k_{\pm}^{\mu}=\left(1,-\,\textrm{sign}(t)m^{i}\right)+O\left(l\right),
\end{equation}
with $m^i=\left(\sin\theta\cos\phi,\sin\theta\sin\phi,\cos\theta\right)$ being the radial unit 3-vector. Finally, using the Clausius relation yields the time derivative of Clausius entropy
\begin{equation}
\label{dS/dt}
\frac{\text{d}S_{C}(t)}{\text{d}t}=\frac{2\pi k_B c}{\hbar}t\int_{\mathcal{S}\left(t\right)}T_{\mu\nu}\left(x\left(t,\theta,\phi\right)\right)k_{\pm}^{\mu}k_{\pm}^{\nu}\text{d}^2\mathcal{A}+O\left(l^4\right)+O\left(\psi\frac{l_P^2a^2}{c^4}\right)+O\left(\frac{1}{a^2}\right).
\end{equation}
This equation (after integration in time) agrees with the semiclassical result~\cite{Baccetti:2013ica,Alonso-Serrano-Liska} up to acceleration-dependent correction terms. However, for $\psi\ne0$, sending the acceleration to infinity would make the sub-leading correction term $O\left(\psi l_P^2a^2/c^4\right)$ dominant (and eventually infinite), breaking the correspondence of both results. For $\psi=0$, no such problem arises, and both results are identical.

While we kept the acceleration, $a$, finite, simply for mathematical convenience, there exist a proposal that quantum gravity in fact limits the maximal attainable acceleration~\cite{Caianiello:1984}. It can be shown that this proposal leads to corrections to Unruh temperature consistent with our modified formula if we set $\psi=2$~\cite{Luciano:2019}. Then, our upper limit on the acceleration, $a\lesssim 1/\sqrt{\psi}=c^2/\sqrt{2}l_P$, agrees with the proposed maximal acceleration up to a factor $1/\sqrt{2}$. In this way, our construction can be made consistent with the maximal acceleration theory.

Integration of the time derivative of $S_{C}$ from bifurcation surface $\mathcal{B}$, at $t=0$, to diamond's future apex $A_f$, at $t=l/c$, yields the total flux of Clausius entropy across the GLCD horizon during its lifetime. To explicitly evaluate the integral, we expand the energy-momentum tensor around the origin of coordinates, $T_{\mu\nu}\left(x\left(t,\theta,\phi\right)\right)=T_{\mu\nu}\left(P\right)+O\left(l\right)$ and use $\int m^{i}\text{d}^2\Omega=0$, $\int m^{i}m^{j}\text{d}^2\Omega=4\pi\delta^{ij}/3$~\cite{Jacobson:2017}. The details of calculations were already discussed by the authors in~\cite{Alonso-Serrano-Liska}. Final expression for the entropy flux results
\begin{equation}
\label{modified_clausius}
\Delta S_{Clausius}=-\frac{8\pi^2 k_Bl^4}{9\hbar c}\left(T_{00}\left(P\right)+\frac{1}{4}T\left(P\right)\right)+O\left(l^5\right)+O\left(\psi\frac{l_P^2a^2}{c^4}\right)+O\left(\frac{1}{a^2}\right).
\end{equation}

The next step in deriving gravitational dynamics is to demand thermodynamic equilibrium, i.e., compensation of the decrease of Clausius entropy by an increase of entanglement entropy $S_{e,q}$ associated with the GLCD horizon. By doing so, we implicitly assume that Clausius entropy is, at least in the leading order in $l$, equivalent to the matter entanglement entropy, The validity of this assumption and details of the comparison were carefully analysed in~\cite{Alonso-Serrano-Liska}. Considering the most general formula for entanglement entropy with a logarithmic correction, its change between times $t=0$ and $t=l/c$ equals
\begin{equation}
\label{entanglement_decrease}
\Delta S_{e,q}=-4\pi\eta l^2+\frac{4\pi\eta l^4}{9}G_{00}\left(P\right)-k_B\mathcal{C}\ln\left(\frac{4\pi\eta l^2-\frac{4\pi\eta l^4}{9}G_{00}\left(P\right)}{\mathcal{A}_0}\right)+O\left(l^5\right).
\end{equation}
Clearly, one obtains non-zero $\Delta S_{e,q}$ even in the cases in which $S_C$ identically equals zero (more starkly, $\Delta S_{e,q}=-4\pi\eta l^2-k_B\mathcal{C}\ln\left(4\pi\eta l^2/\mathcal{A}_0\right)$ in flat spacetime). Therefore, as argued in detail in~\cite{Alonso-Serrano-Liska}, deriving the gravitational equations of motion requires a subtraction of the ``equilibrium state contribution'' corresponding to MSS from $S_{e,q}$, before comparing it with $S_C$ (see also~\cite{Svesko:2017}, where a similar requirement emerges in the context of null cone thermodynamics). The difference yields
\begin{equation}
\label{pure entanglement decrease}
\Delta S_{e,q}-\Delta S_{e,q\,MSS}=\eta\delta\mathcal{A}+k_{B}\mathcal{C}\frac{\delta\mathcal{A}}{\mathcal{A}_{MSS}}-k_B\frac{\mathcal{C}}{2}\left(\frac{\delta\mathcal{A}}{\mathcal{A}_{MSS}}\right)^2+O\left(\left(\delta\mathcal{A}\right)^3\right),
\end{equation}
where
\begin{align}
\nonumber \delta\mathcal{A} &=\frac{4\pi\eta l^4}{9}\left(G_{00}\left(P\right)+\lambda\left(P\right)g_{00}\left(P\right)\right), \\
\nonumber \mathcal{A}_{MSS} &=4\pi l^2+\frac{4\pi l^4}{9}\lambda\left(P\right)g_{00}\left(P\right),
\end{align}
and $\lambda$ is again approximately constant inside the GLCD but in general dependent on $P$ and $l$. Just like in the previous derivation, we can divide $\lambda$ into separate contributions
\begin{equation}
\lambda=\lambda_0+\lambda_q,
\end{equation}
where $\lambda_0=R/4$ is the semiclassical value and $\lambda_q$ contains the quantum corrections proportional to $l_P^2$. In contrast to the MVEH approach, we have no contribution due to non-conformal fields, $\lambda_{nc}$. This is because the Clausius entropy flux depends only on the traceless part of energy-momentum tensor and, therefore, is identical for both conformal and non-conformal fields~\cite{Alonso-Serrano-Liska}.

In total, the following thermodynamic equilibrium condition holds
\begin{align}
& \Delta S_{e}-\Delta S_{e\,MSS}+\Delta S_{C}=0. \label{clausius_balance_1} 
\end{align}
That results in
\begin{align}
& \nonumber \eta\left(S_{00}\left(P\right)-\lambda_q\right)+\frac{k_B\mathcal{C}}{72\pi}\left(S_{00}\left(P\right)-\lambda_q\right)^2+\frac{k_B\mathcal{C}}{4\pi l^2}\left(S_{00}\left(P\right)-\lambda_q\right)\\
& -\frac{2\pi k_B}{\hbar c}\left(T_{00}\left(P\right)+\frac{1}{4}T\left(P\right)\right)+O\left(l\right)+O\left(\psi\frac{l_P^2a^2}{c^4}\right)+O\left(\frac{1}{a^2}\right)=0. \label{clausius_balance_2}
\end{align}
For the same reasons as in the previous subsection, all the remaining $l$-dependent terms must cancel out together. Furthermore, as long as acceleration $a$ remains very large but smaller than $1/\sqrt{\psi}$, it can be set to an arbitrary value and the dependence on $a$ is continuous. Therefore, we can differentiate with respect to it and, by the same reasoning as with $l$, the resulting equation cannot depend on $a$ either. In total, we require
\begin{align}
\nonumber -\eta\lambda_q-\frac{k_B\mathcal{C}}{72\pi}\left(S_{00}\left(P\right)-\lambda_q\right)\lambda_q-\frac{\pi k_B}{2\hbar c}T\left(P\right)\\
+\frac{\mathcal{C}k_B}{4\pi l^2}\left(S_{00}\left(P\right)-\lambda_q\right)+O\left(l\right)+O\left(\psi\frac{l_P^2a^2}{c^4}\right)+O\left(\frac{1}{a^2}\right) &=\Phi\left(P\right), \label{def_Phi}
\end{align}
where $\Phi$ is a scalar function independent of $l$, $a$ and $\psi$ (as all the $\psi$-dependent terms also depend on~$a$).

From here on, the derivation is identical to the one developed in subsection~\ref{MVEH_modified_equations}, without any caveats, so, for simplicity, we skip here the intermediate steps. Let us remind that, as in previous derivation, we fix $\eta=k_B/4l_P^2$ to recover Einstein equations in the limit of $\mathcal{C}\to0$, use independence of the final equations on $n^{\mu}$, invoke EEP to make our result valid throughout the spacetime and determine the remaining unknown scalar ${^{(0)}\Phi}$ from the trace of equations of motion. In the end, we get the following modified gravitational dynamics
\begin{equation}
S_{\mu\nu}-\frac{\mathcal{C}l_{P}^2}{18\pi}S_{\mu\lambda}S^{\lambda}_{\;\:\nu}+\frac{\mathcal{C}l_{P}^2}{72\pi}\left(R_{\kappa\lambda}R^{\kappa\lambda}-\frac{1}{4}R^2\right)g_{\mu\nu}=\frac{8\pi G}{c^4}\left(T_{\mu\nu}-\frac{1}{4}Tg_{\mu\nu}\right).
\end{equation}

\subsection{Comparison of the derivations}

Let us now compare the equations found by both derivations we performed. The form of the final equations of motion is the same, although they involve quantum expectation values in the MVEH approach and classical quantities in the Clausius entropy flux one. The second difference is the numerical factor in front of the $l_{Pl}^2S_{\mu\lambda}S^{\lambda}_{\;\:\nu}$ term, which equals $-\mathcal{C}/30\pi$ and $-\mathcal{C}/18\pi$, respectively. This difference appears because in MVEH approach we performed the variation at fixed volume, while in Clausius flux one we instead considered the entropy changes due to energy-momentum flux across the GLCD's horizon. Since first derivation presented uses a fully quantum definition of entropy while, in second one, we worked with the semiclassical Clausius entropy; a difference in corrections to equations of motion stemming from quantum gravity effects is perhaps to be expected. The difference could be related to the need to hold fixed a quantity known as generalised volume, rather than the usual volume, when dealing with MVEH derivation of equations of motion for modified theories of gravity~\cite{Bueno:2017}. Alternatively, it might imply that the semiclassical equivalence between Clausius and matter entanglement entropy~\cite{Alonso-Serrano-Liska} does not extend to situations where quantum gravitational effects become relevant (although, given the similarity of the equations, both entropies would still have to be closely related). In any case, both derivations yield traceless equations of motion that include a correction term proportional to $l_{Pl}^2\left(S_{\mu\lambda}S^{\lambda}_{\;\:\nu}-R_{\kappa\lambda}R^{\kappa\lambda}g_{\mu\nu}/4+R^2g_{\mu\nu}/16\right)$.

Given the close similarity of both results, we can introduce a general form of the modified equations of motion
\begin{equation}
\label{quantum EoM}
S_{\mu\nu}-Dl_{P}^2S_{\mu\lambda}S^{\lambda}_{\;\:\nu}+\frac{Dl_{P}^2}{4}\left(R_{\kappa\lambda}R^{\kappa\lambda}-\frac{1}{4}R^2\right)g_{\mu\nu}=\frac{8\pi G}{c^4}\left(T_{\mu\nu}-\frac{1}{4}Tg_{\mu\nu}\right),
\end{equation}
where $D=\mathcal{C}/30\pi$ for the equations obtained from MVEH and $D=\mathcal{C}/18\pi$ for the ones derived by the Clausius entropy approach. This difference of factor $5/3$ is insignificant as the coefficient $\mathcal{C}$ in the modified entropy formula is anyway currently unknown. Therefore, in the following, we work with this general form for the quantum phenomenological gravitational equations of motions. Let us now analyse its features in detail.

\section{Interpretation of the modified dynamics}
\label{interpretation}

We introduce this section by noting some general properties of the modified dynamics we derived. Due to being traceless, modified equations of motion do not imply local energy-momentum conservation. Therefore, it needs to be added as an additional assumption. Then, the local conservation condition, $T_{\mu\;\:;\nu}^{\;\:\nu}=0$, implies
\begin{equation}
\label{conservation condition}
\frac{1}{4}R_{;\mu}-Dl_{P}^2\left(S^{\lambda\nu}S_{\mu\lambda}\right)_{;\nu}+\frac{Dl_{P}^2}{2}\left(R^{\kappa\lambda}R_{\kappa\lambda;\mu}-\frac{1}{4}RR_{;\mu}\right)=-\frac{2\pi G}{c^4}T_{;\mu}.
\end{equation}
This condition has relevant implications for the nature of the modified dynamics. Firstly, it cannot be generally solved for $T$. This prevents us from recasting equations of motion in the Einstein-like form that would directly include $T_{\mu\;\:;\nu}^{\;\:\nu}=0$. Secondly, as for all thermodynamically derived gravitational equations of motion, the cosmological constant $\Lambda$ would appear as an arbitrary integration constant in the solution of the conservation condition (we will demonstrate this for the case of a cosmological model in subsection~\ref{cosmology}). Due to these features, our result seems to be a direct generalisation of classical UG, rather than GR. It can be seen that it is not even possible to rewrite the modified equation as a direct generalisation of Einstein equations. Note that the form of quantum corrections is fully determined by the logarithmic term in horizon entanglement entropy. Since many different methods of calculating entropy predict the emergence of such a term due to quantum gravity effects, our conclusions are rather robust. Furthermore, both derivations we presented are completely insensitive to the controversial issue of corrections to Hawking and Unruh temperature. Since both effects are completely kinematic and independent of gravitational dynamics~\cite{Visser:2003}, the fact that their modifications do not enter the modified equations serves as  a consistency check for our derivation.

Both the value and sign of coefficient $\mathcal{C}$ in the modified entropy formula differ in various sources (see subsection~\ref{log_entropy} and references listed there). However, in comparison with the squared Planck length $l_{P}^2\approx2.6\times10^{-70}\,\text{m}^2$, any possible value can be considered to be of the order of unity. Thus, the corrections we find become relevant only when the curvature length scale becomes close to the Planck scale, although it still has to be significantly larger than $l_P$ to view spacetime as a Lorentzian manifold.

We expect that any terms that contain higher than fourth derivatives of the metric or are more than quadratic in curvature tensors will be suppressed by higher powers of $l_P$. This seems to be the case both on dimensional grounds and from the way they appear in the RNC metric expansion. However, the higher derivative terms known from quadratic gravity can be expected to appear at the same order as the corrections we introduce. They are implicitly present on the right hand side of our equations in the expectation value of the energy-momentum tensor~\cite{Wald:1994}. In principle, one might find the higher derivative contributions to the left hand side by a careful analysis of higher order corrections to the area variation. However, these corrections are ambiguous, as they depend on shape deformations of the horizon~\cite{Jacobson:2017}. Without a physically motivated solution of these ambiguities, it is not possible to obtain any insight into how the higher order corrections influence gravitational dynamics. The higher derivative terms that might arise from these corrections are anyway contained with undetermined constants in the energy-momentum tensor expectation value, so their omission does not change the resulting dynamics in any significant way.

Nonetheless, there are good reasons to treat the modifications we propose apart from those of quadratic gravity. Our equations of motion are qualitatively different from those of any local, diffeomorphism invariant theory of gravity. It is most obvious from the, generally unsolvable, condition for local energy-momentum conservation. The difference can be probably traced to the non-locality of logarithmic term in entropy (see~\cite{Solodukhin:2020}, it is also apparent from non-existence of a local expression for entropy density). We intend to explore the related issues in a future work. Furthermore, there are physically reasonable situations in which effects of our theory can be studied apart from those of quadratic gravity. Consider any spacetime with vanishing Weyl tensor and scalar curvature, for example a radiation dominated homogeneous isotropic universe with vanishing cosmological constant. Then quadratic gravity predicts the dynamics known from GR. However, our equations retain correction terms quadratic in the Ricci tensor and, therefore, imply modifications to GR dynamics (similar to the model we discuss in subsection~\ref{cosmology}).

\subsection{Diffeomorphism invariance and the equivalence principle}
\label{Invariant?}

We now turn our attention to the status of three of the cornerstones of GR, full diffeomorphism invariance, local Lorentz invariance and strong equivalence principle, in our approach.

The equations of motion involve only quantities that transform as tensor fields with respect to any diffeomorphism coordinate transformation. Thus, at first glance, one could conclude that the quantum phenomenological gravitational dynamics we introduced retain the full diffeomorphism invariance of GR. However, as we discussed, our result appears to be a generalisation of classical UG rather GR. Actions of UG break the full diffeomorphism invariance by introducing a fiducial volume element or a non-dynamical flat metric~\cite{Barcelo:2014}. Furthermore, semiclassical thermodynamics of spacetime actually seems to point to a variant of UG known as Weyl transverse gravity~\cite{Alonso-Serrano-Liska}, that is explicitly invariant under Weyl transformations (for details of the theory see, e.g.~\cite{Alvarez:2013,Carballo-Rubio:2015,Barcelo:2018}). Weyl invariant theories of gravity often involve higher derivatives of the metric, but Weyl transverse gravity is classically equivalent to GR, replacing the full diffeomorphism invariance. Since the modified equations also include only second derivatives of the metric, it seems likely that breaking diffeomorphism invariance will prove necessary to state the corresponding action.  Nevertheless, a proposal for a fully diffeomorphism invariant version of Weyl transverse gravity does exist~\cite{Jirousek:2019,Hammer:2020}. It modifies the Henneaux-Teitelboim action for UG~\cite{Henneaux:1989}, that retains diffeomorphism invariance by introducing an auxiliary vector density as a new dynamical variable. In conclusion, it can be expected that the modified dynamics do not posses full diffeomorphism invariance, since even their classical limit is only invariant with respect to transverse diffeomorphisms and Weyl transformations (unless dynamical degrees of freedom besides the metric are introduced). Nevertheless, the issue cannot be completely resolved without providing a variational formulation of the modified dynamics.

In this paragraph we discuss some technical concerns regarding the validity of EEP and WEP, and their relation with local Lorentz invariance. The validity of EEP can be restated as local Lorentz invariance of the theory together with weak equivalence principle (WEP), that states: ``Test particles with negligible self-gravity behave, in a gravitational field, independently of their properties.''~\cite{Casola:2015} (see the discussion in~\cite{Casola:2015} for details and possible caveats). UG does not break local Lorentz invariance (Lorentz transformations do not affect determinant of the metric), so the apparently unimodular form of our equations leads to no issues. Let us check the validity of WEP in the modified dynamics, i.e., whether test particles with negligible self-gravity move along geodesics. Since the singular character of point-like particles is problematic in gravitational physics, we will model the test particle as uncharged dust with energy-momentum tensor $T_{\mu\nu}=\rho u_{\mu}u_{\nu}$, where $u^{\mu}$ is the dust velocity. It is easy to check that the local energy-momentum conservation condition, $T_{\mu\;\:;\nu}^{\;\:\nu}=0$, implies geodesic motion of the dust~\cite{Wald:1984}. Thus, the divergence-free energy-momentum tensor implies WEP. In GR, WEP is directly built into the theory, as $T_{\mu\;\:;\nu}^{\;\:\nu}=0$ are four of the ten independent gravitational equations of motion. In most formulations of UG, the divergence-free energy-momentum tensor must be added as an independent assumption  already on the classical level. However, the fully diffeomorphism versions imply it directly. Since the question of full diffeomorphism invariance remains unresolved, we are currently unable to provide a final answer regarding the validity of WEP. On the other hand, \textit{assuming} that the energy-momentum tensor is divergence-free, as is usually done in UG (although there are exceptions~\cite{Josset:2017}), implies the validity of WEP. In our case, it amounts to satisfying the previously stated condition on the trace of energy-momentum tensor. How (and whether) this condition holds in practice probably can be addressed only by studying particular solutions of equations of motion. Once it is satisfied, the modified gravitational equations imply WEP. If EEP violations were to occur even with  WEP satisfied, it would have to be due to the presence of local Lorentz invariance breaking quantum fields (see, e.g.~\cite{Liberati:2013}), a possibility independent of the modified gravitational dynamics we introduced. Therefore, we conclude that, our equations together with the local conservation condition imply the weak equivalence principle. Unless the matter content breaks the Lorentz symmetry even in flat spacetime, the Einstein equivalence principle is then implied as well.

Checking the validity of SEP is a more demanding task. One might try to use the Newtonian limit of modified dynamics. Take a static solution for dust that satisfies the usual Newtonian conditions of weak field and low velocities; we get a modified equation for the gravitational potential $\Phi$\footnote{This form of the Newtonian limit is obtained perturbatively, using a procedure we explain in the following subsection.}
\begin{equation}
\label{Newtonian limit}
\Delta\Phi=4\pi G\rho\left(1-4\pi D\frac{\rho}{\rho_{P}}\right),
\end{equation}
where $\rho$ denotes the matter density and $\rho_P=c^5/G^2\hbar$ is the Planck density. This limit could be understood as implying an effective gravitational constant, $G_{eff}=G\left(1-8\pi D\rho/\rho_P\right)$, dependent on matter content. Then, SEP would be broken, as $G_{eff}$ is not a universal constant. However, the correction term becomes relevant only for very high densities ($\rho_P\approx5.2\times 10^{96}\,\text{kg}\times\text{m}^{-3}$), where the Newtonian approximation completely breaks down. Whenever applicable, the Newtonian limit is equivalent to that of GR. Thus, the Newtonian limit implies no measurable corrections for low energies and velocities, just as expected from quantum gravity corrections that should be relevant only for very high energy densities.

Since the Newtonian limit of modified dynamics is equivalent to that of GR, no violations of SEP appear at low energies. On the other hand, SEP is expected to hold even well beyond the applicability of the Newtonian limit. To confirm or deny its validity in general would require a more involved analysis. A possible method to do so was proposed~\cite{Casola:2014} (the test requires an additional assumption that SEP is equivalent to the combination of local Lorentz invariance and gravitational weak equivalence principle).

\subsection{Application to a simple cosmological model}
\label{cosmology}

To illustrate consequences of the obtained quantum phenomenological gravitational dynamics, we briefly analyse a simple cosmological model. Consider a homogeneous, isotropic, spatially flat universe described by the following metric
\begin{equation}
\text{d}s^2=-c^2\text{d}t^2+a\left(t\right)^2\left(\text{d}r^2+r^2\text{d}\Omega^2\right),
\end{equation}
where $a\left(t\right)$ is the scale factor. For simplicity, we choose a universe filled with dust ($T_{\mu\nu}=\rho\delta^{0}_{\mu}\delta^{0}_{\nu}$). Due to existing symmetries, modified equations of motion yield only one non-trivial condition on the metric, a modified Raychaudhuri equation that can be written as
\begin{equation}
\label{FLRW dynamics 2}
\dot{H}-D\frac{l_{P}^2\dot{H}^2}{c^2}=-4\pi G\rho,
\end{equation}	
where $H=\dot{a}/a$ is the Hubble parameter and the dot denotes the coordinate time derivative. Since the modified equations of motion were obtained from expansion of Bekenstein entropy around its semiclassical value, $S=k_B\mathcal{A}/4l_P^2$, in powers of $l_P^2$, we can likewise consider that the modified Hubble parameter can be obtained as expansion around the classical one, $H_0$
\begin{equation}
H=H_0+l_P^2H_1+O\left(l_P^4\right).
\end{equation}
Given that $\dot{H}_0$ satisfies the standard Raychaudhuri equation, we find
\begin{equation}
\label{FLRW dynamics}
\dot{H}=-4\pi G\rho\left(1-4\pi D\frac{\rho}{\rho_{P}}\right).
\end{equation}
Assuming local energy-momentum conservation implies $\rho=\rho_0/a^3$, with $\rho_0$ being an arbitrary constant with the dimensions of energy density. Substituting for $\rho$ in the previous equation leads to a second order differential equation for $a$
\begin{equation}
\label{equation for a}
\frac{\ddot{a}}{a}-\frac{\dot{a}^2}{a^2}=-4\pi G\frac{\rho_0}{a^3}\left(1-4\pi D\frac{\rho_0}{\rho_{P}a^3}\right).
\end{equation}
If we multiply this equation by $2\dot{a}/a$ and integrate it in time, we obtain a modified Friedmann equation
\begin{equation}
\label{Friedmann_mod}
\left(\frac{\dot{a}}{a}\right)^2=\frac{8\pi G\rho_0}{3a^3}\left(1-\frac{2\pi D\rho_0}{a^3\rho_P}\right)+\tilde{\Lambda},
\end{equation}		
or, in terms of $H$ and $\rho$,
\begin{equation}
\label{Friedmann_mod2}
H^2=\frac{8\pi G\rho}{3}\left(1-\frac{2\pi D\rho}{\rho_P}\right)+\tilde{\Lambda}.
\end{equation}
The arbitrary integration constant $\tilde{\Lambda}$ corresponds to the cosmological term ($\tilde{\Lambda}=\Lambda c^2/3$). This again points to the unimodular nature of the modified dynamics.

Here we encountered no issues with demanding local energy-momentum conservation. However, this is only due to high symmetry of the cosmological model we considered. Indeed, if we take covariant divergence of our equations, the Bianchi identities allow us to rewrite its curvature part in terms of Weyl tensor (plus an extra term we can easily deal with). Since Weyl tensor vanishes in FLRW spacetimes, the local energy-momentum conservation condition is easily satisfied. In general, local energy-momentum conservation may be violated. In other words, if one formulates initial value problem for our equations and specifies divergence-free energy-momentum tensor on the initial Cauchy hypersurface, the time evolution will generically lead to a violation of this condition. This behaviour is connected with the presence of a term proportional to $R_{\mu\lambda}R_{\nu}^{\;\:\lambda}$ (whose divergence can be rewritten in terms of Weyl tensor and a term proportional to $\left(RR_{\mu}^{\;\:\lambda}\right)_{;\lambda}$ via the Bianchi identities). The presence of this term might be connected with the non-locality of logarithmic corrections to entropy~\cite{Solodukhin:2020}. While it remains an open issue at this stage, one possible way to make time evolution generally consistent with local energy-momentum conservation might be along the lines discussed in~\cite{Kegeles:2015}, although perhaps there are other possibilities.

From mathematical point of view, there also exists another solution of the modified Raychaudhuri equation that cannot be found perturbatively. We show in the next that it is not physically relevant. To understand its features, we study the case $\rho=0$ for which the equation can be solved analytically
\begin{equation}
\dot{H}-D\frac{l_{P}^2\dot{H}^2}{c^2}=0.
\end{equation}
Setting $\dot{H}=0$ clearly solves the equation. This is also the solution one finds by using the perturbative method described above. It corresponds to maximally symmetric spacetimes (de Sitter, anti-de Sitter and Minkowski) and introduces no modifications to the classical geometry. However, there also exist a second solution, $\dot{H}=c^2/Dl_P^2$, that cannot be found perturbatively. In this case, the scale factor $a$ equals
\begin{equation}
a=\tilde{a}\exp\left(\frac{c^2}{2Dl_P^2}t^2+\tilde{H}t\right),
\end{equation}
where $\tilde{a}$ is a dimensionless constant and $\tilde{H}$ a constant with dimensions $s^{-1}$. While this solves the modified Raychaudhuri equation, it clearly diverges in the classical limit $l_P\to0$ (or, equivalently, $D\to0$). Since the modified Raychaudhuri equation certainly reduces to the classical one in the same limit, this solution is inconsistent with the nature of corrections we introduced. Therefore, we conclude that, for the case of homogeneous, isotropic universes, perturbatively obtained solutions are the only physically relevant ones (presence of non-zero energy density does not change the basic situation). The same conclusion likely holds even for completely general spacetimes. However, since one then deals with a system of non-linear differential equations, its justification would likely be much more subtle.
	
Also note that, for the perturbative form of the modified Raychaudhuri equation, Picard-Lindelöf theorem for ordinary differential equations ensures existence of a unique solution of the initial value problem (for any $\rho$ independent of $\ddot a$). In contrast, we have seen that there exist two different solutions of the modified Raychaudhuri equation before the perturbative rewrite (although the additional solution is not physically viable). We presume that this insight applies to the general form of the modified equations as well. If the correction terms are perturbatively rewritten as functions of the energy-momentum tensor, the equations will become linear in second derivatives of the metric (assuming minimal coupling and putting aside curvature dependence of its quantum expectation value, which should lead only to $O(l_P^4)$ error). Then, Leray theorem should be sufficient to guarantee uniqueness of the solution of initial value problem.

We proceed by comparing our results with two other approaches to quantum modified cosmology. As expected, for $D>0$, our result strongly resembles equations found in the effective description of loop quantum cosmology (LQC)~\cite{Ashtekar:2015}
\begin{equation}
\label{Friedmann_LQC}
H^2=\frac{8\pi G}{3}\rho\left(1-\frac{\rho}{\rho_{sup}}\right),
\end{equation}
and
\begin{equation}
\label{Raychaudhuri_LQC}
\dot{H}=-4\pi G\left(\rho+p\right)\left(1-2\frac{\rho}{\rho_{sup}}\right).
\end{equation}
Comparison with our equations implies $\rho_{sup}=\rho_{P}/2\pi D$.	Effective dynamics of LQC replaces the Big Bang singularity by a non-singular quantum bounce. Therefore, assuming \mbox{$D>0$}, that is logarithmic corrections to the GLCD entanglement entropy are positive, our modified equations of motion already change the classical gravitational dynamics sufficiently to avoid the cosmological singularity. Up to this point, it seems that the appearance of corrections with $D<0$ would not only keep the singularity but strenghten it. This issue should be addressed in detail in a future work.

Modifications to Friedmann-Lema\^{i}tre-Robertson-Walker (FLRW) universes induced by a modified entropy were already studied~\cite{Awad:2014}. In that case, a thermodynamic derivation of modified Friedmann equations was carried out, taking into account GUP-induced corrections to Bekenstein entropy of the global apparent horizon of a FLRW spacetime. For the case of a spatially flat, dust-filled universe, the following modified Raychaudhuri equation was found
\begin{equation}
\label{GUP FLRW}
\dot{H}-\beta\frac{l_{P}^2H^2\dot{H}}{8c^2}=-4\pi G\rho,
\end{equation}
where $\beta$ is a dimensionless GUP parameter (both $\beta>0$ and $\beta<0$ cases were considered). In the light of the fact that our modified Friedmann equation is perturbatively equivalent to
\begin{equation}
\dot{H}+\frac{3Dl_{P}^2H^2\dot{H}}{2c^2}=-4\pi G\rho,
\end{equation} 
we can see that, upon setting $D=-\beta/12$, both approaches yield the same result. Since, in contrast to our local construction, a thermodynamic derivation based on global features specific to FLRW spacetimes was considered~\cite{Awad:2014}, the equivalence of resulting dynamics is non-trivial and even somewhat surprising. Perhaps this equivalence is specific for the highly symmetric  FLRW geometry, but it might also point to some deeper connection between both approaches.

To conclude this subsection, let us remark that the presence of a quantum bounce can be explained in terms of entropic force acting on the apparent horizon. The entropic force is defined as~\cite{Verlinde:2011}
\begin{equation}
\label{entropic_force}
F_{\mu}\equiv TS_{,\mu}.
\end{equation}
The radius of the apparent horizon obeys $r_A=c/H$. Therefore, the corresponding Kodama-Hayward temperature~\cite{Rocha:2018} and the modified entanglement entropy, respectively, equal
\begin{align}
T&=\frac{\hbar\vert H\vert}{2\pi k_B}\left(1+\frac{\dot{H}}{2H^2}\right) \label{temperature}, \\
S_{e,q}&=\frac{\pi k_Bc^2}{l_P^2H^2}+\mathcal{C}k_B\ln\left(\frac{\pi c^2}{\mathcal{A}_0 l_P^2H^2}\right) \label{entropy_mod}.
\end{align}
The only non-trivial component of entropic force is $F_0$, for which it holds
\begin{equation}
F_{0}=-\text{sign}\left(H\right)\frac{\hbar c}{l_P^2}\left(\frac{\dot{H}}{H^2}+\pi\mathcal{C}\frac{l_P^2}{c^2}\dot{H}\right)\left(1+\frac{\dot{H}}{2H^2}\right).
\end{equation}
Expressing $H^2$ and $\dot{H}$ from the modified Friedmann and Raychaudhuri equations yields
\begin{equation}
F_{0}=\text{sign}\left(H\right)\frac{3\hbar c}{8l_P^2}\frac{\left(1-\frac{4\pi D\rho}{\rho_P}\right)\left(1+\frac{4\pi D\rho}{\rho_P}\right)}{1-\frac{2\pi D\rho}{\rho_P}}\left(\frac{1}{1-\frac{2\pi D\rho}{\rho_P}}+6\pi^2\mathcal{C}\frac{\rho}{\rho_P}\right),
\end{equation}
that, for an expanding universe and $D>0$ is negative for $\rho>\rho_P/4\pi D=\rho_{sup}/2$ and, likewise, positive for the same densities in a contracting universe. Hence, in the very early phase of expansion, entropic force acting on the apparent horizon is repulsive. Understanding entropic force in relation with the gravitational interaction, its repulsiveness explains why a regular bounce occurs instead of the Big Bang singularity.
	
One can read the modified Raychaudhuri equation as a standard one with an effective pressure term $-4\pi Dc^2\rho^2/\rho_P$. Then, this term violates the null and dominant energy condition precisely in the regime in which the entropic force is repulsive (for an expanding universe). However, the weak energy condition, $\rho\ge0$, still holds.

\section{Discussion}
\label{discussion}

In this paper, we propose novel and general quantum phenomenological gravitational dynamics. The only assumption made about the effects of quantum gravity is the presence of a logarithmic correction term in the entanglement entropy associated with spherical local causal horizons. Given the number of conceptually different methods of calculating entropy that predict the presence of a logarithmic term (at the very least in the case of black hole entropy), our conclusions are quite robust and relevant to many approaches to quantum gravity, e.g. LQG, string theory, path integral quantum gravity, AdS/CFT correspondence and GUP phenomenology.

Specifically, we provided two independent derivations of quantum phenomenological equations of motion for gravitational dynamics from thermodynamics of geodesic local causal diamonds. The result represents a direct generalisation of the classical equations of motion of unimodular gravity. This is consistent with a previous work of the authors~\cite{Alonso-Serrano-Liska}, that argued for unimodular character of classical gravitational dynamics derived from thermodynamics. Here, we found that quantum corrections lead to a generalisation of unimodular equations of motion that cannot be restated as generalised Einstein equations. Then, equivalence of UG and GR holds on the level of classical dynamics, is broken as predicted. Therefore, we clearly showed that, due to quantum gravity entering into the game, thermodynamics of spacetime imply unimodular gravitational dynamics distinct from any generalisation of GR.

Several attempts to exploit logarithmic corrections to entropy in order to obtain some insight into quantum modifications of the gravitational dynamics were made in recent years. They were based on the entropic force model of gravity~\cite{Verlinde:2011} applied either to a global static background~\cite{Majumder:2013,Feng:2016,Kibaroglu:2019} or to FLRW spacetimes~\cite{Awad:2014,Salah:2016}. However, the similarities between these works and our derivation are only superficial. The most fundamental difference lies in the fact that we use a local construction applicable in any general spacetime. Furthermore, in contrast with the previous attempts, we provide general gravitational equations of motion which do not include any undetermined functions.

The approach to phenomenological quantum gravity introduced in the present paper still requires further development. In the future research, we plan to provide a detailed analysis of constraints on parameter $D$, and, especially, find an action which implies the modified equations of motion we obtained. Finding the action should settle the crucial questions of diffeomorphism invariance and local energy-momentum conservation. Furthermore, it would be interesting to check whether exponentially growing modes appear in linearised perturbations of the equations, indicating instability. We expect that the solutions will be stable as long as the perturbative approach to dynamics outlined in subsection~\ref{cosmology} is justified. However, an explicit calculation might lead to some unexpected findings. In addition, we are already exploring modifications of explicit solutions known from general relativity. This might provide some insight into the effects of quantum gravity corrections on gravitational systems.

Let us remark that our results also strengthen the semiclassical equivalence between Clausius and entanglement entropies~\cite{Alonso-Serrano-Liska}. While the exact equivalence possibly breaks on the quantum level, both entropies remain strongly related. Their precise relation should be analysed carefully in a future work.

In summary, we have presented foundations for a new phenomenological perspective on the low energy quantum gravitational dynamics. Whether and how will this perspective affect our understanding of quantum gravity remains to be seen.

\acknowledgments 

The authors want to acknowledge Luis Garay and Matt Visser for relevant discussions and for reading earlier drafts of the paper. Also, the authors acknowledge discussions with Jos\'{e} Senovilla and with participants of \mbox{CARRAMPLAS} workshop. A. A-S. also wants to acknowledge discussion with Miguel Zumalacarregui that was at the seed of this work. Finally, we want to thank the anonymous referee for pointing out interesting issues to introduce in the discussion.

A. A-S. is funded by the Alexander von 	Humboldt  Foundation. A.A-S work is also partially supported by the Project. No. MINECO FIS2017-86497-C2-2-P from Spain. This article is based upon work from COST Action CA15117 ``Cosmology and Astrophysics Network for Theoretical Advances and Training Actions (CANTATA)'', supported by COST (European Cooperation in Science and Technology).

\end{document}